# AUTOMATIC ENERGY SAVING (AES) MODELTO BOOST UBIQUITOUS WIRELESS SENSOR NETWORKS (WSNs)

[1]Abdul Razaque
Department of Computer Science and Engineering
University of Bridgeport
CT-06604, USA
[1]arazaque@bridgeport.edu

[2]Khaled Elleithy
Department of Computer Science and Engineering
University of Bridgeport
CT-06604, USA
[2]elleithy@bridgeport.edu

**Abstract:** We deploy BT node (sensor) that offers passive and active sensing capability to save energy. BT node works in passive mode for outdoor communication and active for indoor communication. The BT node is supported with novel automatic energy saving (AES) mathematical model to decide either modes. It provides robust and faster communication with less energy consumption. To validate this approach, network simulator-2 (ns2) simulation is used to simulate the behavior of network with the supporting mathematical model. The main objective of this research is to remotely access different types of servers, laptops, desktops and other static and moving objects. This prototype is initially deployed to control MSCS [13] & [14] from remote place through mobile devices. The prototype can further be enhanced to handle several objects simultaneously consuming less energy and resources.

## I. INTRODUCTION

The technological progress in size of microprocessors has made a significant advancement for ambient intelligence (AmI) [1]. AmI is the fastest growing segment attracting the people around the world [5], [31], [32]. AmI nourishes from many well organized fields of computing and engineering. It also combines several professions through many application domains, e.g., health, education, security and social care. Many objects are now embedded with computing power like home appliances and portable devices (e.g., microwave ovens, programmable washing machines, robotic hovering machines, mobile phones and PDAs). These devices help and guide us to and from our homes (e.g., fuel consumption, GPS navigation and car suspension) [2]. AmI involves compact power that is adapted to achieve specific tasks. This prevalent accessibility of resources builds the technological layer for understanding of AmI [6], [8].

Information and communications technologies (ICT) have highly been accepted as part of introducing new cost-effective solutions to decrease the cost of pedagogical activities and healthcare. For example, the Ubiquitous intelligence health home that is equipped with AmI to support people in their homes. While this notion had some problems to be fully understood in the past, but due to emerging technologies and incredible progress in low-power electronics and sensor technologies supported with faster wireless network have facilitated the human life. Robust heterogeneous wireless sensor network systems can be organized for controlling mobility of objects and logistics [30]. These developments have led to the introduction of small-sized sensors that are capable of monitoring the constraints of humans as well as living environment [3]. The objective of AmI with use of sensors is to provide better living standard [27].

The deployment of mobile devices as sensor nodes provides more flexibility to interact with objects in any environment [4]. These technologies not only improve the quality of education and health of people in their places, but also provide the fastest way of communication to interact with devices all over the world [9]. The use of mobile devices in wireless sensor network provides more flexibility, intelligence and adaptability to interact with devices dynamically in any environment [11]. This makes it possible to deploy mobile phones not only as terminal nodes, but also as remote controller for several devices. With the deployment of mobile infrastructures, larger area can be covered as compare with stationary infrastructure using same number of sensors [10]. Meanwhile, wireless networks face many challenging issues including unreliable communications, limited energy, storage resources, inadequate computing and harsh environments [29]. In this paper, we introduce a novel paradigm that involves mobile phones and sensors. In particular, our goal is to introduce faster and robust wireless sensor network to facilitate controlling remotely available servers and different devices. Furthermore, the paradigm provides remote accessibility with minimum energy consumption.

## II. RELATED WORK

Controlling remote servers using mobile devices is one of the highly challenging issues because of scalability, interoperatibility, limited services and security of mobile devices. The access of one or multiple servers from remote places saves resources and fosters better communication. The mobile devices supported with sensors make the task quicker and smarter but from other side, sensors consume more energy. No matured paradigm is found in the literature that supports devices being remotely controlled using the sensors and mobile devices while selecting a robust and efficient path. The salient features of most related work are discussed in following studies. Until now, adhoc solution has been deployed in home automation with support of current technologies to fulfill the requirements of users. Java-based home automation system using the Internet was introduced in [22]. The authors manage some digital input and output lines which are connected with the home appliances. Internet based wireless home automation system for multifunctional devices have been introduced in [26] to provide remote access. The authors in [22] and [26] used radio frequency link and simple management protocol to handle these devices.

Home automation solution for indoor ambient intelligence (AmI) has been implemented in [23]. Authors have used gateway and local control panels to maintain the security system for the house. Security involves with communication and performs several entities at the same time. The IP based communication platform has been developed to interact with security company and security staff. Authors in [24] introduced building automation systems to control house appliances with support of typical services and standard applications models. The paper also focused on BACnet, EIB/KNX and Lonworks as open systems in building automation domain. Three-level working model was implanted inside automation pyramid that reduced convolution of individual level and kept levels transparent.

### III. Proposed Architecture

Our proposed architecture consists of two types of devices: mobile phone and sensors BTnode rev3 (Bluetooth-enabled). The mobile phone is initially used to control a single server but it can further be upgraded to control several types of servers and devices. The BTnode rev3 is a self-directed prototyping platform based on microcontroller and Bluetooth radio. It supports distributed sensor networks, wireless communication and ad-hoc networks. It comprises of microcontroller, separate radio and ATmega 128. The radio of BTnode rev3 supports two radios: first is low power chipcon CC1000 reserved for ISM-band and works like Berkeley MICA2 mote. It also supports to establish multi-hop networks.

Second radio is Zeevo ZV4002 to support Bluetooth module. The proposed work covers two types of scenarios. The first scenario is purely a testbed consisting of 3 exterior wireless sensor (EWS)[ BT node rev3] nodes: 1 end device node, 1 Boarder (base station) node and 1 programmable serial interface (PSI) board, digital addressable lighting interface controller (DALIC), one server, active Badge Location (ABL) System, 2 interior wireless sensors(IWS) [BT node rev3] and mobile phone as shown in Figure.1.

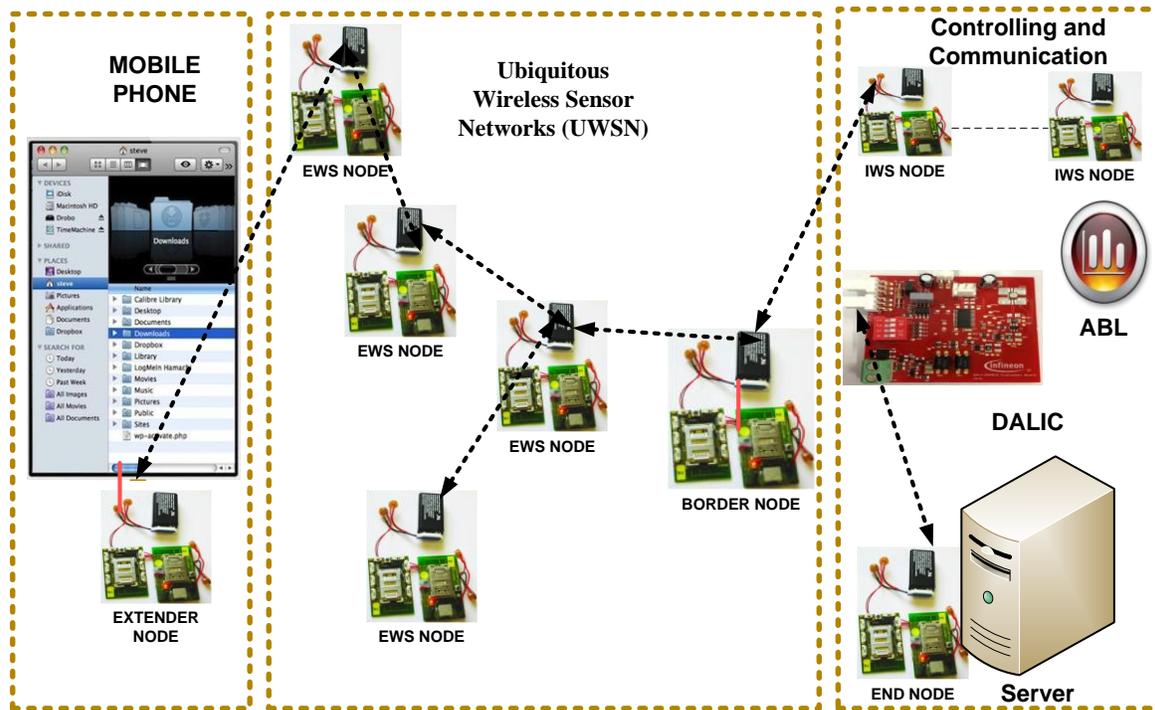

**Figure 1: Ubiquitous wireless sensor network to control Server**

PSI node is connected with mobile phone, DWS nodes are positioned in the outdoor environment of the building, end node is connected with DALIC and server. All of the sensors use radio frequency communication devices to communicate with each other in wireless network. They collectively communicate to turn on/off the server from remote distance, where the boarder node functions as a data hub while end node performs its task for passing the signal to server.Referring to Figure.1, Board of WSN and the DALIC are connected with the server through standard RS232 serial communication port to facilitate the utilization of the ubiquitous communication to manipulate and control process to conserve unused energy. One of the key aspects of testbed simulation is to obtain and manage the working process intelligently through UWSN to provide ample interaction between mobile phone and sensors in both outdoor and indoor environment. We have developed a Java-based interface and application-level framework that allows suitable reuse of sensor-specific code. It separates high-level application logically from underlying sensor driver as shown in Figure 2.

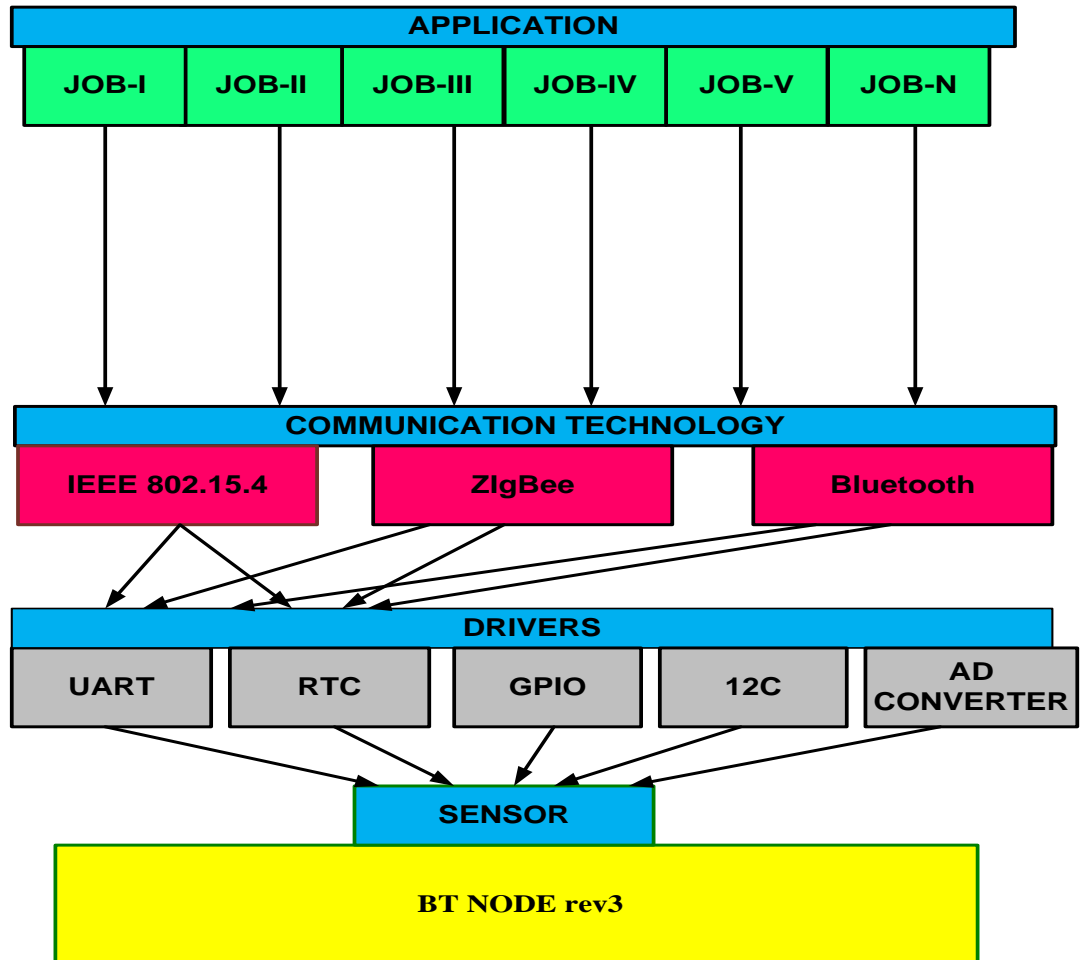

**Figure 2: Application-level framework to reuse the sensor-specific code**

The framework simplifies an application development by generating a single interface that manages virtually all kinds of external and built-in sensors. It reduces the size of the code required to access sensors. PSI board controls sensing with minimum impact on the device's energy utilization. We attach PSI board to back side of cell-phone to build integrated sensing podium. We use the module to replace the use of battery panel that is integrated with PSI board. PSI interfaces with mobile phone's SD card slot, and in resulting it does not increase physically the size of the phone, and even not directly bang the functionality of the mobile phone. It is also designed for extensibility in order to add new classes of applications. We also monitor the daily activity patterns of the mobile phone while controlling the servers.

The sensors provide organized services, where each sensor comprises of a large set of ready to process data. We have used ABL System for finding position of server inside a building. ABL emits unique infrared code after every 10 seconds that is standard time but in our case, we use time 500 millisecond that makes faster detection process. Signals emitted by ABL are collected via IWS sensors around the building. On basis of collected information position, of server is determined. If we control several objects from remote places then we will just attach wireless transmitter to every object to be located. In addition, the matrix of receiving elements with ultrasonic detectors will be deployed on the ceiling of room. This will help to calculate the position of the transmitter using multilateration technique given in [25]. We conclude that our testbed UWSN system is highly sensitive, contextualized, interconnected responsive, transparent and smart enough to control remotely available server with less consumption of energy.

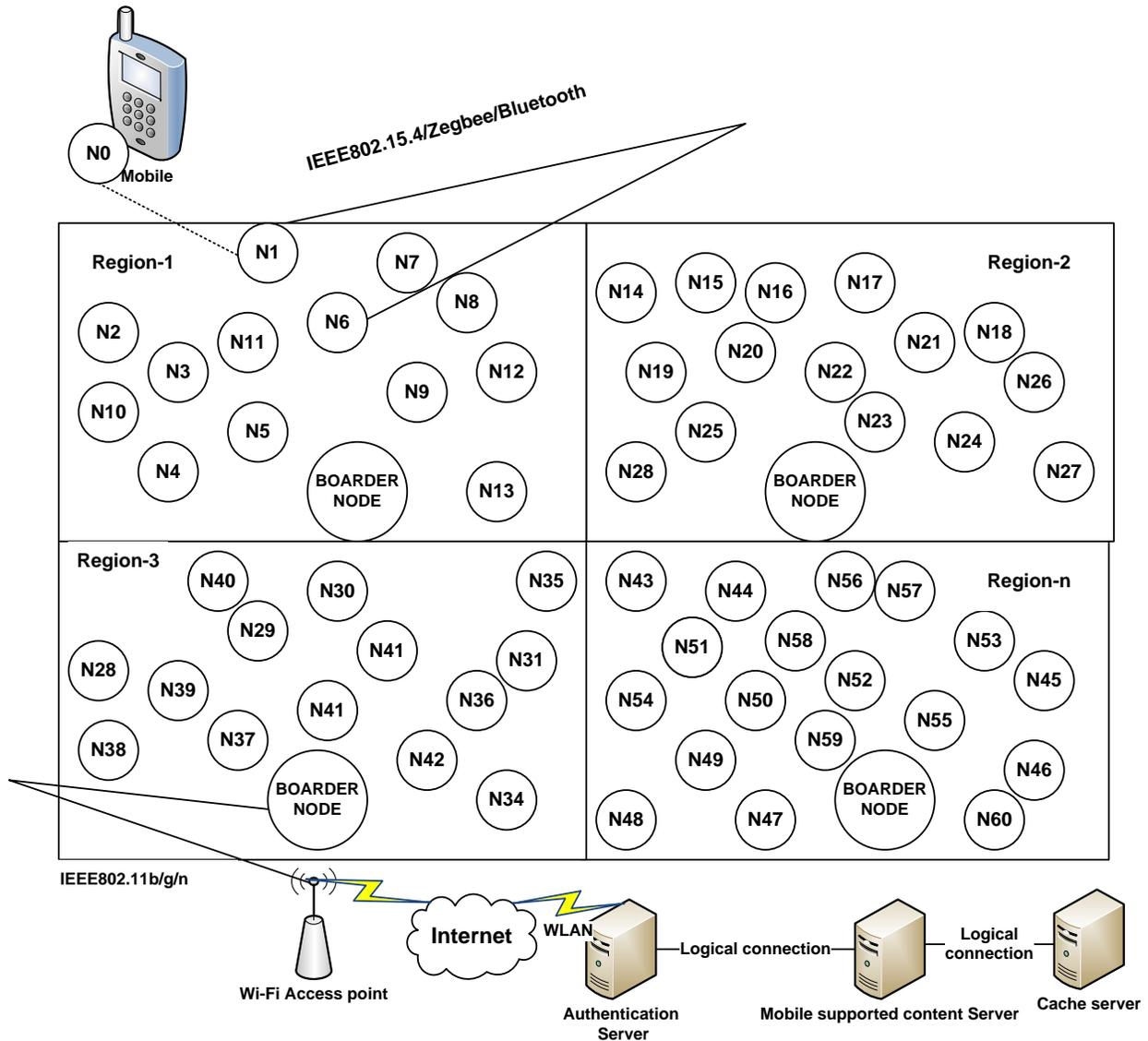

**Figure 3: Proposed wireless sensor network**

This multi-featured network supports to all BT node sensors and different product of mobile families supported with Zigbee/IEEE 802.15.4 and IEEE 802.11 b/g/n standards. Sensors select shortest efficient path on basis of used mathematical model that choose reliable path for sending the data. In addition, on the sensor network, only path-finder sensors are active and remaining sensors are always at sleep state as shown in Figure 3. The advantage of sensor network is the distribution into different regions. Each region has one boundary node that coordinates with the boundary nodes of other regions. This process is also done with AES mathematical model for consuming and gaining energy in active and passive modes. The boundary sensor is the only one that is active and plays a role in communicating with boundary nodes or other nodes of the adjacent regions. The collective working process of the mathematical models, explained later, makes the network highly intelligent to carry the commands of the mobile devices in a faster way to remotely control the available servers and other devices.

## IV. Automatic energy saving (AES) model

It this section we introduce the Automatic Energy Saving (AES) model. Assume the number of sensors is N, which are deployed to detect the presence or absence of indoor and outdoor environment (IOE). Sensor $_K$ collects information pertaining to IOE and makes the appropriate decision based on the collected information. $D_i$ ($D_i$ =1 for deciding presence of indoor environment (IE) and otherwise $D_i$ =0, it means outdoor environment (OE). The detection process is based on the maximized probability of detection (PD), with respect to constraints on probability of unknown environment (UE), and by Neyman-Pearson Lemma [16],[19],[20]. Thus, the IOE environment is identified. With help of blinking observations, Neyman-Pearson Lemma explains optimal detector design of detector but we use this to map for IOE environment. In other words, the sensor detects environment with realization of random variables K, UE and PD. Since the probability of unknown environment (UE) is random variable, which explains constraint of optimized problem in form of UE= α is no more applicable. One important selection for statistical optimized quantity is the expected value of UE and PD. Hence, we maximize the expected value of probability to detect UE, with respect to constraints of expected value of probability [28].

Max E(PD)

E(UE) = α

$$E(UE) = \sum_{k=0}^{\infty} PK\,(K=k)UE^k\,(\beta^k)$$

$$E(PD) = \sum_{k=0}^{\infty} PK\,(K=k)PD^k(\beta^k)$$

Hence,

$UE^k$ and $PD^k$ are linked by relative operating characteristi (ROC) curve, we assume the functional form of relationship is according to $PD^k = f(UE^k)$.

Thus

$$E\,(UE) \sum_{k=0}^{\infty} PK\,(K=k)UE^k\,(\beta^k) \tag{1}$$

$$E(PD) = \sum_{k=0}^{\infty} PK\,(K=k)\,f(UE^k(\beta^k)) \tag{2}$$

The Problem of optimization of constraint can be defined as follows:

Max E(PD)

E (UE) = α

$0 \leq UE^k \leq 1$, k = { 0, 1, 2,…,}

Here,

UE is vector of realization of unknown environment

$UE = [\;UE^1\,(\beta^1)\quad UE^2\,(\beta^2)\quad …]$

In realistic environment, the UE cannot be infinite, since probability of getting many number of samples of environment, P ( k=k) during finite T, reaches 0 as K → ∞, for any avialable environment.

Lagrangian multiplier method can be used to resolove this issues as follows:

$$F = E\,(PD) + \zeta\,[E\{UE\} - \alpha] = \sum_{k=0}^{\infty} PK\,(K=k)PD^k\,(\beta^k) + \zeta \sum_{k=0}^{\infty} PK\,(K=k)\,UE^k(\beta^k) - \alpha$$

Distinguishing with respect to $UE^k$, and equating to 0, So we obtain:

$$\frac{dPD^k\,(\beta^k)}{d\,UE^k\,(\beta^k)} = -\zeta = \gamma \quad k = 0, 1, …\,\infty \tag{3}$$

This can be further expressed as:

$$E\,(UE) \sum_{k=0}^{K=\infty} PK\,(K=k)UE^k\,(\beta^k) = \alpha \tag{4}$$

K is sensor node of N that is case of finite capacity environment, then equation (3) represents N equations in $UE^k, \gamma$, and combine with equation (4). Thuse, we can resolve N +1 by using $UE^k, \gamma$, where k ={ 0, 1, 2, 3…, N}. We can obtain the threshhold of UE using inverse of relationship of $UE^k$ $\beta^k$. Therefore, we can obtain unknown environment but we have also to find indoor and outdoor envirnment.

We are using following probabilties for indoor and out out door environment.

$UE_i$ = P ($D_i$ =0 | IOE outdoor), $\beta i$ = P ($D_i 0$ = 0| IOE outdoor), $PD_i$ = P ($D_i$ = 1| IOE indoor), $\gamma_i$ = P ($D_i 0$ =1| IOE indoor).

Assume that detection of environment is independent made by sensor, the UE probability of obtained decision from ith sensor is given by following equation:

$$\beta = UE\ (1 - Pc1) + (1 - UE)Pc0 \tag{5}$$

Here Pc1(Pc0) is probability of IOE. When sensor ith transmits bit for finding "0" and "1". For IOE, Pc1 = Pc0 = Pc
Therefore, (5) can be simplified as follows:

$$\beta = UE\ + (1 - 2UE)Pc \tag{6}$$

The probability of detection the IOE depends on receiving decision, γ can be obtained from (6) and replacing UE with PD. FOR IOE

$$\gamma = PD\ + (1 - 2PD)Pc \tag{7}$$

This helps to determine the IOE, on basis of decision, BT node automatically works either active or passive mode. This automatic process of detection causes the saving energy.

Where, we replace γ with IOE.

$IOE = PD\ + (1 - 2PD)Pc$

$\gamma\ and\ IOE$ are showing the nature of environment, $Di = (1 - 2PD)Pc$; we substitute $(1 - 2PD)Pc$ with Di. We get:

$IOE = PD\ +\ Di$

$$PD = IOE - Di \tag{8}$$

If we obtain the value of Di, repace probability detection (PD) with its substitute values.

$$IOE = \sum_{k=0}^{\infty} PK\ (K = k)\ f(UE^k (\beta^k)) + Di$$

Now rearrange and get Di.

$$Di = IOE - \sum_{k=0}^{\infty} PK\ (K = k)\ f(UE^k (\beta^k)) + Di \tag{9}$$

If we obtain the value Di =1, it means passive mode is initiated and sensor $ith$ perserves the energy. Passive mode is supported by energy of sun. Di=0 gives the sign of active mode and BT node consumes the energy in this mode, which obtains through passive mode. If Di ≥ 1 || Di ≤ 0, shows that environment is unknown and sensor $ith$ does not work and goes to sleep position in order to save the energy.

We can obtain energy perserving as follows:

$$EN(X) = \sum_{i=0}^{N} E\ (i)\gamma(ai) \tag{10}$$

$$EN(Y) = \sum_{j=0}^{N} E(j)\partial(aj) \tag{11}$$

Where EN(X) and EN(Y) denote total energy used by two different networks. E(i,j) indicates the energy used by node i and j during transmission.

$$\begin{matrix} 0 & if\ aij\ =\ 0 \\ 1 & otherwise \end{matrix} \tag{12}$$

Assume that aij is different from 0, only if node j receives E(i,j) when node i transmits. This gives appropriate minimum level $E_{min}$ i.e. if

$$E(i,j) > (1 +_{min} rij^{n)}\ E\min \tag{13}$$

We deduce from the equation (13), it consume less energy during the process.

BT sensor node follows the energy saving integration method during the passive process.

We here show numerical time integrators that causes of preserving energy P(e), we begin by assuming an x-point quadrature formula with nodes $N_i$. The required weight of ai is obtained through Lagrange basis polynomials in interruption that is shown as follows:

$$\lim(\tau) = \prod_{j=1, j \neq i}^{x} \frac{\tau - Nj}{Ni - Nj}\ ,\quad ai = \int_{0}^{1} \lim(\tau) d\tau \tag{14}$$

Let $a_1, a_2, a_3,…a_x$ be different real numbers (usually 0 ≤ Ni ≤1) for which $a_i$ ≠ for all i. we use polynomial $p(d_0)$ for satisfying the degree.

$$p(do) = xo \tag{15}$$

$$p(d0 + Nej) = A(p(d0 + Nej) \int_0^1 \frac{\delta y}{\delta x} \nabla S \, (p(d0 + \tau s)) dx \tag{16}$$

The quadrature formula with nodes $N_i$ and weights $a_i$ decrease integrator to specific collection of methods. We use polynomial degree $2_x - 1$, thus Gauss points $N_i$ that is equal to 0 and shifted with Lagrange polynomial specific collection for A(x). This treats arguments in A(x) and $\nabla S\,(x)$ with different way that is considered as partitioned numerical method. The solution of these methods depends on specific factorization of vector filed.

Assume, If A(x) = A is constant matrix, let (1, 1) be Hamiltonian system, thus it becomes energy saving integrator. This is proof that sensor node also consumes minimum amount of energy during passive mood.

## V. SIMULATION SETUP AND ANALYSIS OF RESULT

We compared AES with TRAMA [7], S.MAC [21], TMAC [12], MMAC [17] and CMAC [15, 18]. TRAMA and MMAC represent schedule-based MAC protocols, whereas SMAC, TMAC and CMAC embody contention-based MAC protocols. TRAMA and MMAC are scheduled based MAC protocols, which outperform SMAC and TMAC in this experiment. When , mobility of devices is employed, packets dramatically decrease in SMAC, TMAC, TRAMA, CMAC and MMAC whereas AES model demonstrates a minimal decrease. Our objective is to save energy because it is considered as most significant performance feature for WSNs. For our experimental simulation setup, we use ns-2.35-RC7. The wireless sensor network is distributed into different regions as illustrated in Figure 3 to make the sensors more convenient to collect information quicker. We have already discussed the role of the boundary node as anchor point (AP) or head node.

We have set one boundary node in each region. The boundary node forwards the collected information of its region to the next region. In our case, it is not necessary that the boundary node may always coordinate with only boundary node of another region but it can forward the gathered information at 1-hop destination either boundary node or any active node. We have simulated realistic scenario that is a real test of WSN. We have deployed 105 sensors within network area of 200m × 200 m. Area is divided into 40m x 40m regions. Sensors are randomly located within each region. The sink in this scenario is located at (140, 60). The bandwidth of node is 50 Kbps and the maximum power consumption for each sensor was set 160 mW, 12 mW and 0.5 mW for communication, sensing and idle modes respectively but in our case, there is no idle mode. Sensors either go to active or sleep mode. Each sensor is capable of broadcasting the data at power intensity ranging from -20 dBm to 12 dBm.

The total simulation time is 40 minutes and there is no pause time during the simulation but we set 30 seconds for initialization phase at start of the simulation. During this phase, only sensors onto sink remain active and the remaining sensors of all regions go into power saving mode automatically. The results shown in this section are average of 12 simulation runs.

### A. Consumption and Saving Energy

The primary goal of this research is to consume less energy during the communication between the mobile devices and the remotely placed servers and several devices. Maximizing the life time of the WSN, AES helps to consume minimum energy. We have established 30 connections and transferred 30 GB data at the rate off 377 *10$^{-6}$ seconds/byte. The total numbers of sensors have consumed 151 joule energy in our case, whereas TRAMA, CMAC, SMAC and TMAC have consumed between 300 to 500 joule shown in Figure 4. It is demonstrated through the simulation that AES model has saved 87.2% energy as shown in Figure 5. We have used 14 transmitters in this scenario and compared the energy consumption per node for TRAMA, CMAC, TMAC, SMAC and AES. Based on the results of this simulation, each node consumes less energy in AES as compared with other MAC protocols as shown in Figure 6. In addition, AES also uses less total duty cycles shown in Figure. Transmission of the generated data in the whole network can be obtained as:

$$N(a) = 2\pi \int_{r+t/2}^{R} \lambda x dx * Td \tag{1}$$

Generation speed of data in $N(a)$ is obtained as:

$$N(a) = 2\pi \int_{r+t/2}^{R} \lambda x dx * Ds \tag{2}$$

'Ds' is speed of generated data in the whole network that is constant K.
Total energy consumption speed in region-N can be calculated as:

$$N(a) = 2\pi \int_{r+t/2}^{R} \lambda x dx * Vj \tag{3}$$

Thus, total consumption of energy Tc(E) is deduced to combine (9), (10) and (11), we get:

$$Tc(E) = 2\pi \int_{r+t/2}^{R} \lambda x dx * Vj + 2\pi \int_{r+t/2}^{R} \lambda x dx * Ds + 2\pi \int_{r+t/2}^{R} \lambda x dx * Td \tag{4}$$

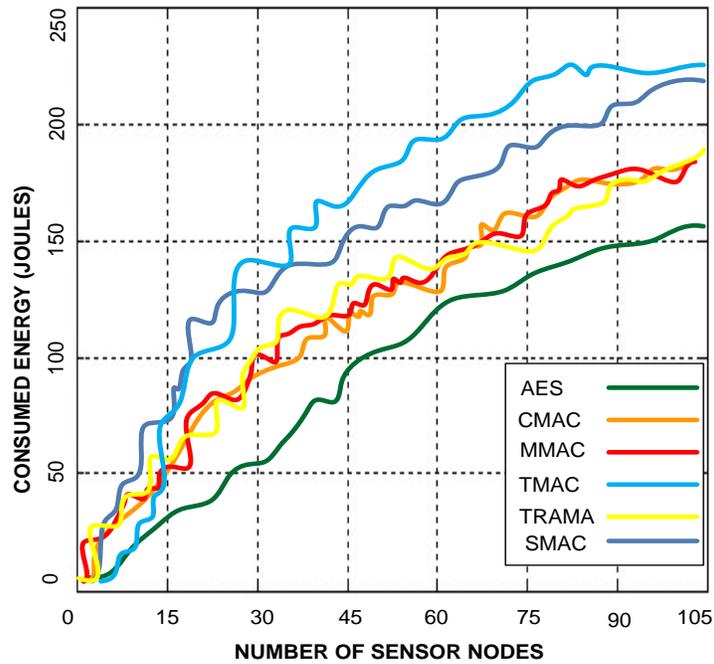

**Figure 4: Energy consumption at different number of sensors**

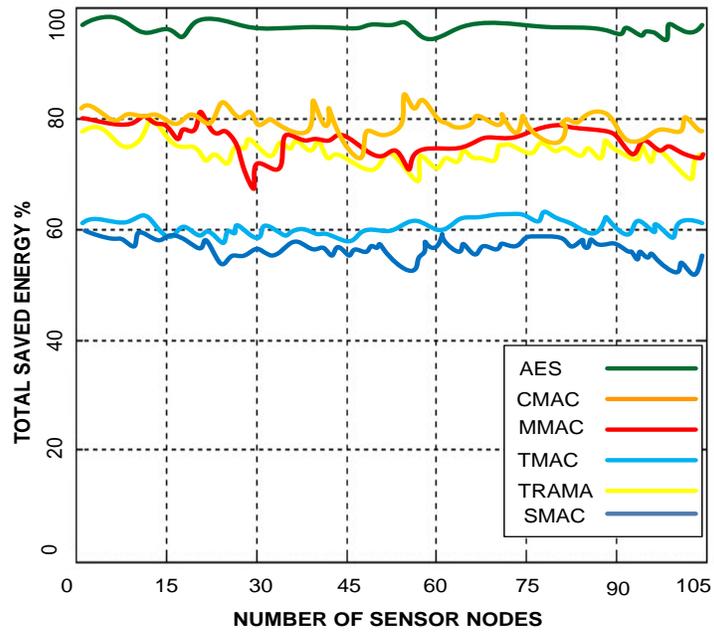

**Figure 5.Total Energy saving for number of sensors**

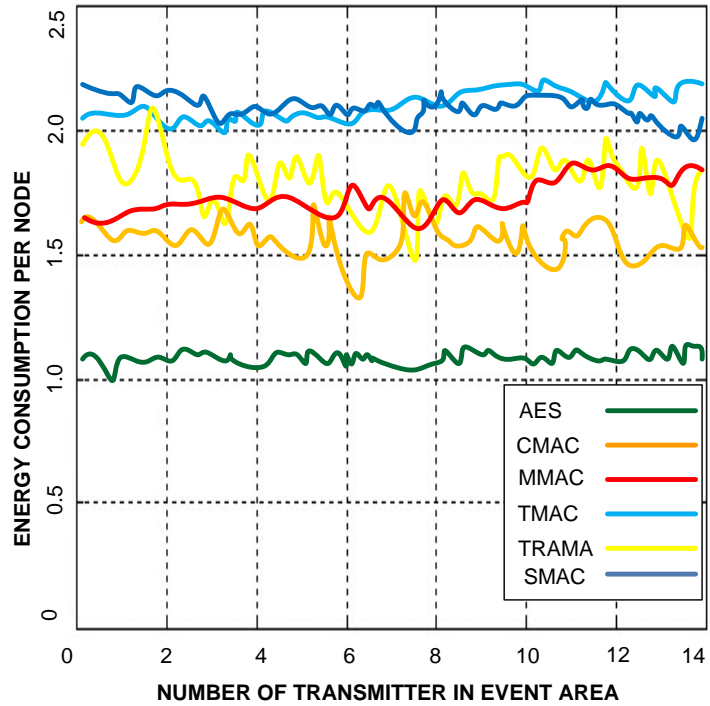

**Figure 6: Consumption of energy at different transmitters**

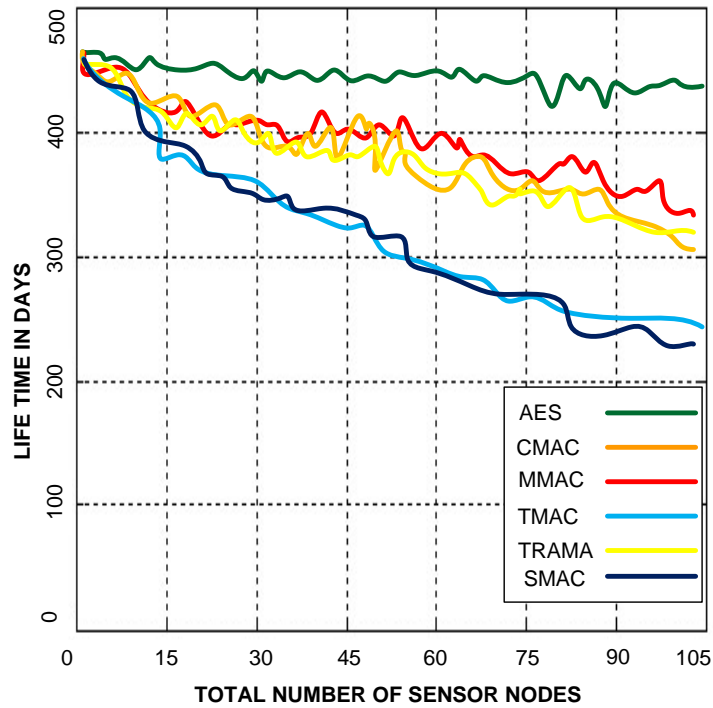

**Figure 7: Life time of sensors over WSNs at different approaches**

## VI. CONCLUSION

In this paper, we have introduced an automatic energy saving (AES) model that senses the environment to work on either passive or active mode using BT node sensors for saving energy. AES model enforces the sensors to work on

1- hop destination that also causes energy saving. Furthermore, the network is divided into a number of regions N. Each boundary node is responsible to communicate with other regions for consuming minimum energy of its region. The advantage of this model is to activate only one sensor in the whole region. When the sensor finishes its task, it automatically goes to a sleep mode. To validate the proposed model in WSN, we implemented the proposed model in a testbed and ns2.35-RC7.

Based on the simulation results, we demonstrated that the proposed research have saved maximum amount of energy compared with similar techniques. In addition, we were able to control the remotely available servers and devices while consuming the minimum energy. The testbed simulation was implemented to control the servers and ns2 simulation was used to demonstarte the robustness of paradigm.

In the future, we are planning to implement the applications of this research to control house automation devices, office devices and several static and moving devices while consuming minimum energy.